\documentclass[aps,twocolumn,reprint,prl,preprintnumbers,superscriptaddress]{revtex4-1}
\usepackage{amsmath,amssymb,amsthm,mathtools,enumitem,multirow,bm,graphicx,float,array,pgf,soul,bm,comment,booktabs,adjustbox,csquotes,mathdots,hyperref}
\usepackage{braket,mleftright}
\usepackage[para]{threeparttable}
\usepackage[all]{xy}
\usepackage[normalem]{ulem}

\usepackage{comment}

\newcommand{\cA}{\mathcal{A}}

\newcommand{\GMo}{GM^{(4)}|_{a = \frac{1}{3}}}

\newcommand{\cW}{\mathcal{W}}
\newcommand{\cM}{\mathcal{M}}

\usepackage{xcolor}
\definecolor{amethyst}{rgb}{0.54, 0.17, 0.89}
\definecolor{coral}{rgb}{1.0, 0.3, 0.4}

\begin{document}

\title{Constraints on four-party entanglement in holography}

\author{Vijay Balasubramanian}
\email[\texttt{vijay@physics.upenn.edu}]{}
\affiliation{Department of Physics and Astronomy, University of Pennsylvania, Philadelphia, PA 19104, U.S.A}
\affiliation{Theoretische Natuurkunde, Vrije Universiteit Brussel, 
%and International Solvay Institutes, Pleinlaan 2, 
B-1050  Brussels, Belgium}
\affiliation{International Centre for Theoretical Sciences-TIFR,
Shivakote, Hesaraghatta Hobli, Bengaluru North 560089, India}

\author{William~K.L.~Chan}
\email[\texttt{chanwill@sas.upenn.edu}]{}
\affiliation{Department of Physics and Astronomy, University of Pennsylvania, Philadelphia, PA 19104, U.S.A}

\author{Monica~Jinwoo~Kang}
\email[\texttt{monicak@tamu.edu}]{}
\affiliation{Mitchell Institute for Fundamental Physics and Astronomy, Texas A\&M University, College Station, TX 77843, U.S.A.}

\author{Chitraang~Murdia}
\email[\texttt{murdia@sas.upenn.edu}]{}
\affiliation{Department of Physics and Astronomy, University of Pennsylvania, Philadelphia, PA 19104, U.S.A}

\author{Simon F. Ross}
\email[\texttt{s.f.ross@durham.ac.uk}]{}
\affiliation{Centre for Particle Theory, Department of Mathematical Sciences, Durham University, Durham DH1 3LE, U.K.}

\date{\today}

\begin{abstract}
We show that in pure time-reflection-symmetric holographic states several known four-party entanglement signals vanish unless the triple information $I_3$ is non-zero.  In this sense, our results show that $I_3$ is the strongest known signal of the presence of quadripartite entanglement.  Additionally, $I_3$ quantitatively bounds all four-party entanglement signals built from the multi-entropy. However, the residual entropy $Q_4$, also a measure of four-party entanglement, is not bounded by $I_3$, although $I_3=0$ does imply $Q_4=0$ for holographic states (except on a set of measure zero for which $Q_4$ is ill-defined).
%\CM{Is $Q_4$ well-defined on this set? If not we could say that "(except on a set of measure zero for which $Q_4$ is ill-defined)".}
%These bounds establish new constraints on the entanglement structure of holographic states. 
%\CM{Not sure if we need the last sentence.}
%%%%%
%are bounded by the triple information $I_3$. Thus, $I_3$ is a stronger signal of four-party entanglement than the multi-entropies. By contrast, we show that the residual entropy $Q_4$ is not bounded by $I_3$, although $I_3=0$ implies $Q_4=0$ except for a measure zero set of holographic states. These bounds give new constraints on the entanglement structure of holographic states. 
%%%%%
%Remarkably, however, we show that whenever $I_3$ vanishes, a large class of four-party entanglement signals also vanish. These results identify $I_3$ as an organizing measure of four-party entanglement and suggest a hierarchical structure of multipartite entanglement in holographic states.
\end{abstract}

\maketitle

\section{Introduction}\label{sec:intro} 

Understanding the nature of entanglement between subsystems of a quantum system is a fundamental question in quantum information theory, quantum many-body physics, and holography. While bipartite entanglement can be characterized using the von Neumann entropy, entanglement between multiple parties is much less understood. In holography, entanglement in the boundary state is related to the bulk geometry. Bipartite entanglement as measured by the entanglement entropy is related to the area of minimal surfaces in the bulk through the Ryu-Takayanagi (RT) formula and its generalizations
\cite{Ryu:2006bv, Hubeny:2007xt}. Similar geometric descriptions for various signals of multiparty entanglement have been proposed, see e.g. \cite{Dutta:2019gen,Akers:2019gcv,Gadde:2022cqi,Balasubramanian:2024ysu,Harper:2020wad, Bao:2021vyq,Jain:2022csf,Ju:2025eyn}. 
These correspondences between entanglement structures and geometric features of the bulk allow us to explore patterns of  entanglement in holographic states. Geometric relations between the areas of RT surfaces lead to a set of constraints on the state formalized via the holographic entropy cone \cite{Bao:2015bfa, Marolf:2017shp, Avis:2021xnz, Hernandez-Cuenca:2022pst,Czech:2023xed,Czech:2024rco}.

Relations between quantities sensitive to multiparty entanglement have begun to be explored \cite{Balasubramanian:2025hxg}, showing that pure time-symmetric holographic states can never possess only GHZ-like entanglement. Here, we obtain new relations between signals of four-party entanglement that hold, like those in the holographic entropy cone,  for holographic states, and not for general quantum states. 
Multiparty entanglement signals have also been explored in \cite{Bao:2025psl, Ahn:2025bdm, Naskar:2026zka}.

Signals of four-party entanglement are quantities calculated for a pure state $\ket{\psi}_{ABCD}$ of a system with four parts which vanish if the state only involves entanglement between non-trivial subsets of the parts. We consider three kinds of signals in this work. 

First, we have the triple information \cite{Hayden:2011ag}
\begin{equation}
    I_3 = S_A + S_B + S_C + S_D - S_{AB} - S_{AC} - S_{BC} ,
\end{equation}
which was shown to be a signal of four-party entanglement in \cite{Balasubramanian:2014hda}. 
The triple information does not have a definite sign in a general quantum state, but it is non-positive for holographic states \cite{Hayden:2011ag}. 

Second, we have signals constructed from the multi-entropy $S^{(q)}$ introduced in \cite{Gadde:2022cqi} (see also \cite{Penington:2022dhr}). The three-party signal was first constructed in \cite{Gadde:2023zzj,Harper:2024ker}, and the extension to $n$-party signals was developed in \cite{Iizuka:2025ioc, Iizuka:2025caq}, where these signals were referred to as genuine multi-entropies. Our interest is in the four-party genuine multi-entropy,
\begin{multline}
\label{eq:GM4_def}
    \GMo 
    = S^{(4)}(A:B:C:D) - \frac{1}{3} \sum S^{(3)} \\
    + \frac{1}{3} \left( S_{AB} + S_{BC} + S_{AC} \right) ,
\end{multline}
where the sum is taken over all distinct tripartitions of $ABCD$. The notation $a = \frac{1}{3}$ denotes a specific choice in the general definition given in \cite{Iizuka:2025ioc}; other choices correspond to adding a multiple of the triple information. 
A systematic discussion of multi-party entanglement signals was recently given in \cite{Gadde:2026msg}, which identified additional four-party signals constructed from the multi-entropies. The space of these signals is spanned by $I_3$, $\cM_{AB|CD}$, $\cM_{AC|BD}$, and $\cM_{AD|BC}$.
Here,
\begin{multline} 
    \cM_{AB|CD} = S^{(4)}(A:B:C:D) + S_{AB} \\
    - S^{(3)}(AB:C:D) - S^{(3)}(A:B:CD)  ,
\end{multline}
and $\cM_{AC|BD}$, $\cM_{AD|BC}$ are defined similarly by permuting the parties. 
The genuine multi-entropy in \eqref{eq:GM4_def} can be expressed in terms of these quantities as
\begin{equation}
\label{eq:GM4_cM}
    \GMo = \frac{1}{3} \left( \cM_{AB|CD} + \cM_{AC|BD} + \cM_{AD|BC} \right) .
\end{equation}
The definition of the multi-entropy from \cite{Gadde:2022cqi} is somewhat complicated, so we will not review it here. In holography, it is conjectured to be given by the area of certain minimal brane webs \cite{Gadde:2022cqi}, as we review in the next section.   
 
Third, we consider the residual entropy, $Q_4$, defined using a double canonical purification \cite{Balasubramanian:2024ysu}. This is also a signal of four-party entanglement, and is non-negative for holographic states. As it is constructed by canonical purification, it is determined holographically by the area of entanglement wedge cross-sections, as we review in the next section. We will also introduce a new quantity $\widetilde{Q}_4$, defined only holographically, which satisfies $\widetilde{Q}_4 \geq Q_4$.

These different signals are all sensitive to the presence of four-party entanglement, but they are not obviously related. This is not surprising -- we do not expect  a simple notion of the ``amount'' of four-party entanglement in a state. The structure of higher-party entanglement is complicated, and can generally take different forms. 
This is already seen in simple qubit systems \cite{Dur:2000zz,Coecke:2010jha, Hein:2004zjp, Verstraete:2002gqj} and should be even more true as we move to the large-dimensional quantum systems dual to holographic spacetimes. 
It is therefore surprising that we find that these signals satisfy non-trivial relations. 

In this paper, we show that in any holographic system with time-reflection symmetry, the four-party signals constructed from multi-entropies are bounded by $I_3$:
\begin{equation} \label{Mbound}
    0 \leq \cM_{AB|CD} \leq - I_3 . 
\end{equation}
Using \eqref{eq:GM4_cM}, it follows that $\GMo$ satisfies the same bounds. As with the bound in \cite{Balasubramanian:2025hxg}, these relations are true only in holographic systems. 
As a non-holographic counterexample, consider the four-party GHZ state $\frac{1}{\sqrt{2}} (\ket{0000}+\ket{1111})$, which has $I_3 = \ln 2$ \cite{Cui:2018dyq}.
Moreover, $S = \ln 2$, $S^{(3)} = 2 \ln 2$ \cite{Gadde:2022cqi}, and  $S^{(4)} = 3 \ln 2$ \cite{Iizuka:2025caq}, so $\cM_{AB|CD} =0$.
Hence, the four-party GHZ state violates the  bound in \eqref{Mbound}. 

These bounds provide new constraints on the structure of multiparty entanglement in holographic states, namely the triple information quantitatively bounds all measures of four-party entanglement built out of the multi-entropy.  Additionally, we demonstrate that  
\begin{equation}
I_3=0 \quad \implies \quad \widetilde{Q}_4=0 \quad \implies \quad Q_4 = 0 \, ,
\end{equation}
for holographic states, excluding a set of measure zero.  This result formalizes an observation from \cite{Balasubramanian:2024ysu} that there were no known examples where $Q_4$ was non-zero and $I_3$ was zero.   Thus, the triple information must be non-zero for any of the known signals of four-party entanglement to be non-vanishing; in this sense $I_3$ is the strongest known signal of the presence of quadripartite entanglement. On the other hand, we will provide an explicit example in vacuum AdS$_3$ to show that $I_3$ does \emph{not} quantitatively bound the magnitudes $Q_4$ or $\widetilde{Q}_4$ in holographic states, when these signals are nonvanishing.

The remainder of this paper is organized as follows. We first review the holographic definitions and properties of the relevant multiparty entanglement signals. We then present proofs of the bounds in \eqref{Mbound}. Next, we give the counter-example to show that no analogous bound can exist for $Q_4$ and $\widetilde{Q}_4$.  We also demonstrate that if $I_3$ vanishes in an open region of parameter space then $\widetilde{Q}_4$ must also vanish, and hence $Q_4$ too. We conclude with a discussion of implications and possible extensions and other directions for future work.

\section{Signals of multipary entanglement}
\label{subsec:holography}

We begin by reviewing the holographic construction of multipartite entanglement signals, introducing the relevant minimal surfaces. We restrict our discussion to geometries with time reflection symmetry, and compute entanglement at the moment of time symmetry, so the holographic duals of the entanglement quantities are given by minimal surfaces that live on the time reflection symmetric spatial slice. For a boundary region $A$, the entanglement entropy is given by the RT formula \cite{Ryu:2006bv}
\begin{equation}
\label{eq:RT}
	S_A = \frac{1}{4G} \cA[\Gamma_A],
\end{equation}
where $\Gamma_A$ is the minimal bulk surface homologous to $A$.
The region between the RT surface $\Gamma_A$ and the boundary region $A$ in the spatial slice is an initial data surface for the entanglement wedge $EW(A)$ -- the holographic dual for the reduced density matrix $\rho_A$ \cite{Czech:2012bh,Headrick:2014cta, Dong:2016eik, Jafferis:2015del}.
The triple information is then expressed in terms of the areas of the appropriate RT surfaces.
It has been proved that $I_3 \leq 0$ in holography; this property is known as the monogamy of mutual information \cite{Hayden:2011ag}.

We assume, following \cite{Gadde:2022cqi}, that the multi-entropies are given holographically by areas of minimal brane-webs.  Brane-webs are certain collections of surfaces in the bulk spatial slice which can be connected or disconnected, and can also include junctions. 
The brane-web computing the multi-entropy is the minimal candidate that separates each region from the others, i.e., it contains subwebs homologous to each of the boundary regions. 
Given three boundary regions $A$, $B$ and $C$, the  tripartite multi-entropy is 
\begin{equation}
\label{eq:hol-tripartite-multi-entropy}
	S^{(3)}(A:B:C) = \frac{1}{4G} \cA[\cW_{A:B:C}] .
\end{equation}
with $\cW_{A:B:C}$ being the minimal brane-web that separates the three regions. 
Similarly, $S^{(4)}(A:B:C:D)$ is given by the area of the minimal brane-web $\cW_{A:B:C:D}$. We will assume that this proposal is valid, although there is a subtlety in the $n \to 1$ limit in the R\'enyi entropies used to derive it \footnote{In general, the continuation from integer $n$ to $n \to 1$ is a subtle question. For the multi-entropy, \cite{Penington:2022dhr} points out that the assumption that the dominant bulk saddle is replica symmetric fails for $n=3$ in the AdS$_3$/CFT$_2$ case. Further discussion of replica symmetry for bulk saddles in the AdS$_3$/CFT$_2$ case is given in \cite{Gadde:2023zzj}. The situation is similar to that in \cite{Belin:2013dva}, where a phase transition in the ordinary Rényi entropy was found as a function of $n$. Consequently, the R\'enyi entropy need not be analytic for all $n$, but only in a neighborhood of $n=1$. Given that a family of replica-symmetric saddles exists for all $n$ and is dominant at sufficiently small $n$ (namely, $n=2$), it is reasonable to similarly conjecture that the replica-symmetric saddles here determine the behavior in the $n \to 1$ limit, although a formal proof would be challenging. Recently,  \cite{Akella:2026bci} explored the replica symmetry breaking in random tensor network models.}.

Thus, quantities like $\cM_{AB|CD}$ are computed using the areas of relevant brane-webs and RT surfaces. 
Using these surfaces, \cite{Iizuka:2025caq} showed that $\GMo \geq 0$ in certain holographic settings where $A,B,C,D$ are connected boundary regions. 
We will later show holographic positivity more generally. 

\begin{figure}
    \centering
    \includegraphics[width=0.5\linewidth]{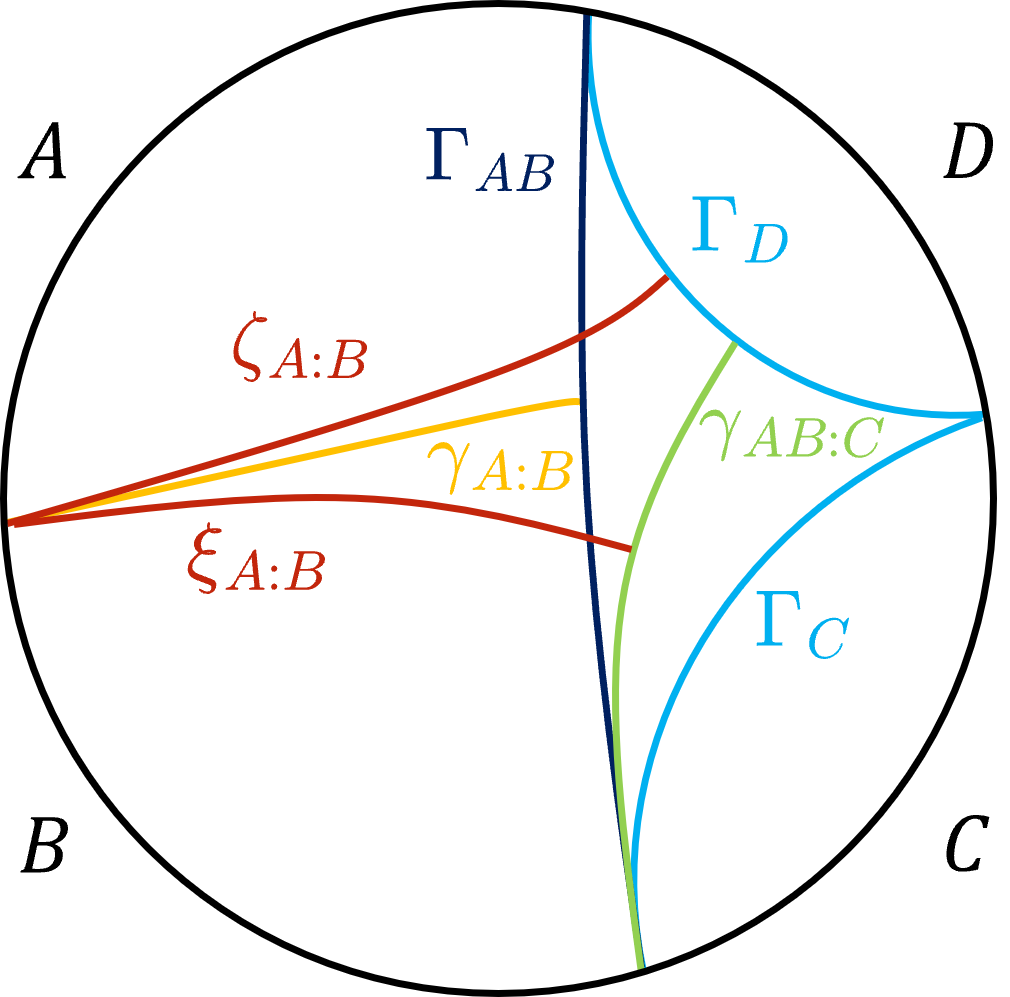}
    \caption{The minimal surfaces used to calculate the reflected entropy $S_R(A:B)$, the residual entropy $Q^{(4)}(A:B)$, and the alternate quantity $\widetilde{Q}^{(4)}(A:B)$.}
    \label{fig:surfaces}
\end{figure}

To describe the holographic computation for residual entropy $Q_4$, we first introduce the reflected entropy. The reduced density matrix $\rho_{AB}$ dual to a bulk entanglement wedge $EW(AB)$ has a canonical purification $\ket{\sqrt{\rho_{AB}}}$, which is a state in $\mathcal H_A \otimes \mathcal H_B \otimes \mathcal H_A^* \otimes \mathcal H_B^*$. Tracing over $B B^*$, we define the reflected entropy as the von Neumann entropy, 
\begin{equation}
    S_R(A:B) = S_{AA^*}(\ket{\sqrt{\rho_{AB}} }).
\end{equation}
It is given holographically by \cite{Dutta:2019gen}
\begin{equation}
    S_R(A:B) = \frac{2}{4G} \cA[\gamma_{A:B}], 
\end{equation}
where $\gamma_{A:B}$ is the  entanglement wedge cross-section -- the minimal surface separating $A$ and $B$ in $EW(AB)$.
Fig.~\ref{fig:surfaces} shows an example in the vacuum AdS$_3$ setting. 

The residual entropy $Q_4$ is given holographically by 
\begin{equation}
    \label{eq:Q4}
    Q_4(A:B) = \frac{1}{4G} \left( \cA[\xi_{A:B}] - \cA[\gamma_{A:B}] \right), 
\end{equation}
where $\xi_{A:B}$ is the minimal surface separating $A$ from $B$ in the portion of $EW(ABC)$ bounded by the entanglement wedge cross-section $\gamma_{AB:C}$ \cite{Balasubramanian:2024ysu}. Note that this differs from the original definition by an overall factor of $\frac{1}{4}$. 
Finally, we introduce an alternate holographic quantity
\begin{equation}
    \label{eq:Q4-tilde}
    \widetilde{Q}_4(A:B) = \frac{1}{4G} \left( \cA[\zeta_{A:B}] - \cA[\gamma_{A:B}] \right),
\end{equation}
where $\zeta_{A:B}$ is the minimal surface that separates $A$ and $B$ within $EW(ABC) \cap EW(ABD)$. The $\xi_{A:B}$ and $\zeta_{A:B}$ surfaces are also illustrated in Fig.~\ref{fig:surfaces}.
Since we do not have a definition for $\widetilde{Q}_4$ outside holography, we do not know if it is itself a four-party entanglement signal. However, it is useful because it is easier to compute and provides an upper bound on $Q_4$ as we show below.

First note that $\gamma_{AB:C}$ is completely contained within $EW(ABC) \cap EW(ABD)$; it lies in $EW(ABC)$ by definition, and by entanglement wedge nesting in the canonical purification it must lie outside of $EW(C)$ and hence inside $EW(ABD)$.
If the surface $\zeta_{A:B}$ does not cross $\gamma_{AB:C}$, it is a candidate for $\xi_{A:B}$. If it does cross $\gamma_{AB:C}$, the portion up to the crossing is a candidate for $\xi_{A:B}$. In either case, $\cA[\zeta_{A:B}] \geq \cA[\xi_{A:B}]$, and hence 
\begin{equation}
\label{eq:Q4-tilde-geq-Q4}
    \widetilde{Q}_4(A:B) \geq Q_4(A:B) \geq 0.
\end{equation}

\section{Bounds on multi-entropy signals}
\label{sec:proofs}

\begin{figure}
    \centering
    \includegraphics[width=\linewidth]{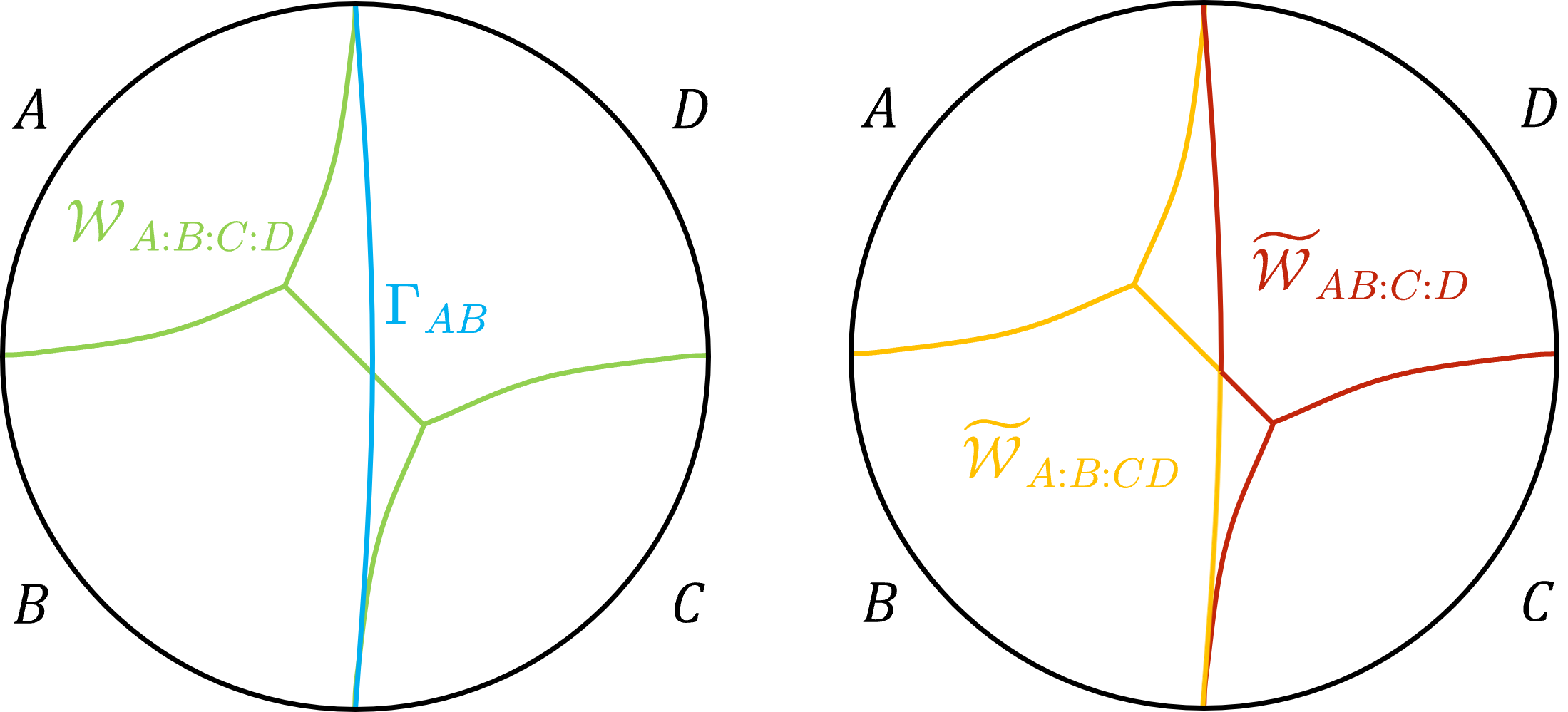}
    \caption{The brane-web $\mathcal{W}_{A:B:C:D}$ (green) and $\Gamma_{AB}$ (blue) shown on the left can be split into two disjoint pieces (right) $\widetilde{\mathcal{W}}_{A:B:CD}$ and $\widetilde{\mathcal{W}}_{AB:C:D}$ depicted in yellow and red, which are larger than the minimal brane-webs $\mathcal{W}_{A:B:CD}$ and $\mathcal{W}_{AB:C:D}$ respectively. }
    \label{fig:mproof}
\end{figure}

We now turn to our main results which provide new constraints on the multi-entropy signals in holography. 
First, we prove the non-negativity of $\cM_{AB|CD}$. 
Consider the joint brane-web given by the four-party brane-web $\mathcal{W}_{A:B:C:D}$ and the RT surface $\Gamma_{AB}$; a representative example is shown in the left panel of Fig.~\ref{fig:mproof}.
This joint brane-web contains at least one surface that separates $A$ from $B$ and $C$ from $D$, and at least two surfaces that separate any other pair of boundary subregions. 
Thus, the joint brane-web can be partitioned into two distinct sub-webs, as illustrated in the right panel of Fig.\ref{fig:mproof} -- one that separates $A$, $B$, and $CD$, and another that separates $AB$, $C$, and $D$. 
The former, $\widetilde{\cW}_{A:B:CD}$, consists of the segments closest to $A$ and $B$, while the latter, $\widetilde{\cW}_{AB:C:D}$, consists of the segments closest to $C$ and $D$. 
As the joint brane-web contains two surfaces separating $AB$ from $CD$, these two sub-webs are either non-overlapping or the overlapping segment is present in both $\mathcal{W}_{A:B:C:D}$ and $\Gamma_{AB}$. Since $\mathcal{W}_{A:B:CD}$ and $\mathcal{W}_{AB:C:D}$ are the minimal brane-webs, it follows that
\begin{multline}
    \cA[\cW_{A:B:C:D}] + \cA[\Gamma_{AB}] \\
    \geq \cA[\cW_{A:B:CD}] + \cA[\cW_{AB:C:D}], 
\end{multline}
and hence, $\cM_{AB|CD} \geq 0$. Using \eqref{eq:GM4_cM}, this implies $\GMo \geq 0$, extending the claim of \cite{Iizuka:2025caq} to arbitrary pure time-reflection-symmetric holographic states.

\begin{figure}
    \centering
    \includegraphics[width=\linewidth]{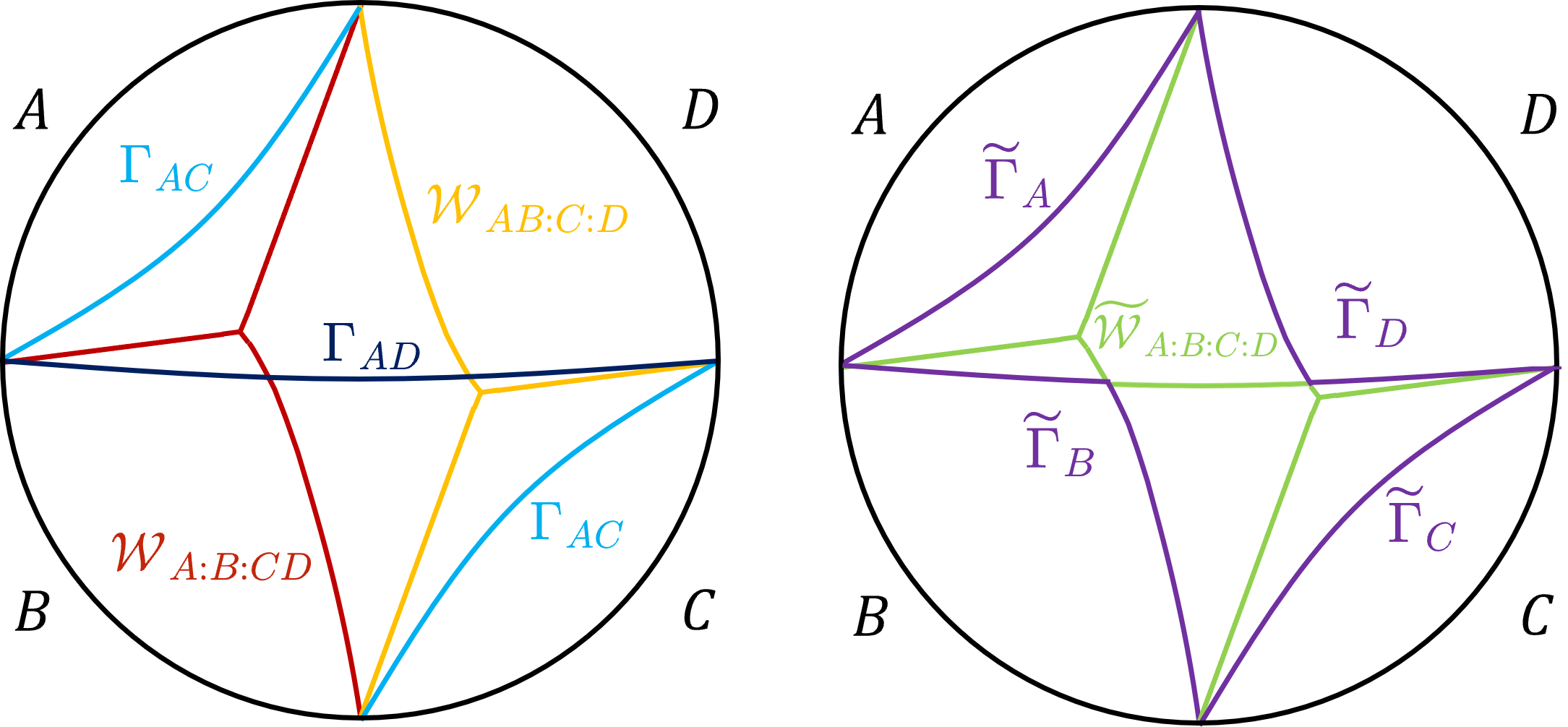}
    \caption{A possible set of surfaces corresponding to the RHS of \eqref{eq:claim1} (left). On the right, in dark blue are the surfaces closest to each boundary subregion. What remains in teal forms a brane web $\widetilde{\mathcal{W}}_{A:B:C:D}$ that separates each subregion from every other subregion.}
    \label{fig:proof1}
\end{figure}

Next, we show that 
\begin{multline}
\label{eq:claim1}
    \cA[\cW_{AB:C:D}] + \cA[\cW_{A:B:CD}]
    + \cA[\Gamma_{AC}] + \cA[\Gamma_{AD}] \\
    \geq  \cA[\Gamma_A] + \cA[\Gamma_B] + \cA[\Gamma_C] + \cA[\Gamma_D] \\
    + \cA[\cW_{A:B:C:D}] .
\end{multline}
Consider the joint brane-web given by the surfaces corresponding to the left side of this equation; the left panel of Fig.~\ref{fig:proof1} shows a representative example.
Each pair of boundary subregions is separated by at least three surfaces contained in this joint brane-web. For instance, $A$ and $B$ are separated at least once by each of $\cW_{A:B:CD}$, $\Gamma_{AC}$, and $\Gamma_{AD}$. 
We now decompose the joint brane-web into five distinct pieces as shown in the right panel of Fig.~\ref{fig:proof1}. 
The first piece, $\widetilde{\Gamma}_A$, is comprised of the segments closest to boundary subregion $A$. Since this is a candidate for the RT surface of $A$, we have $\cA[\widetilde{\Gamma}_A] \geq \cA[\Gamma_A]$. 
Similarly, we can obtain $\widetilde{\Gamma}_B$, $\widetilde{\Gamma}_C$, $\widetilde{\Gamma}_D$ whose areas must be bigger than or equal to the analogous RT surfaces. 
Finally, we are left with $\widetilde{\cW}_{A:B:C:D}$ which still separates all boundary subregions from each other, because removing the $\widetilde{\Gamma}$'s gets rid of only two surfaces that separate any pair of boundary subregions. Evidently, $\cA[\widetilde{\cW}_{A:B:C:D}] \geq \cA[\cW_{A:B:C:D}]$. 
Thus, we can decompose the surfaces corresponding to the left side of \eqref{eq:claim1} into five distinct pieces that have greater or equal areas as the surfaces on the right side of the equation. This concludes our proof.

Adding $S_{AB}$ to both sides of \eqref{eq:claim1} and rearranging, we obtain the upper bound
\begin{equation}
    \label{eq:M-partition-upper-bound}
     \cM_{AB|CD} \leq  -I_3.
\end{equation}
By permuting the subregions, we can get the analogous inequalities $\cM_{AC|BD}\leq -I_3$ and $\cM_{AD|BC} \leq -I_3$.
Using \eqref{eq:GM4_cM}, it follows that $\GMo \leq -I_3$.

\section{Constraints on Residual Entropy}

\begin{figure}
    \centering
    \includegraphics[width=\linewidth]{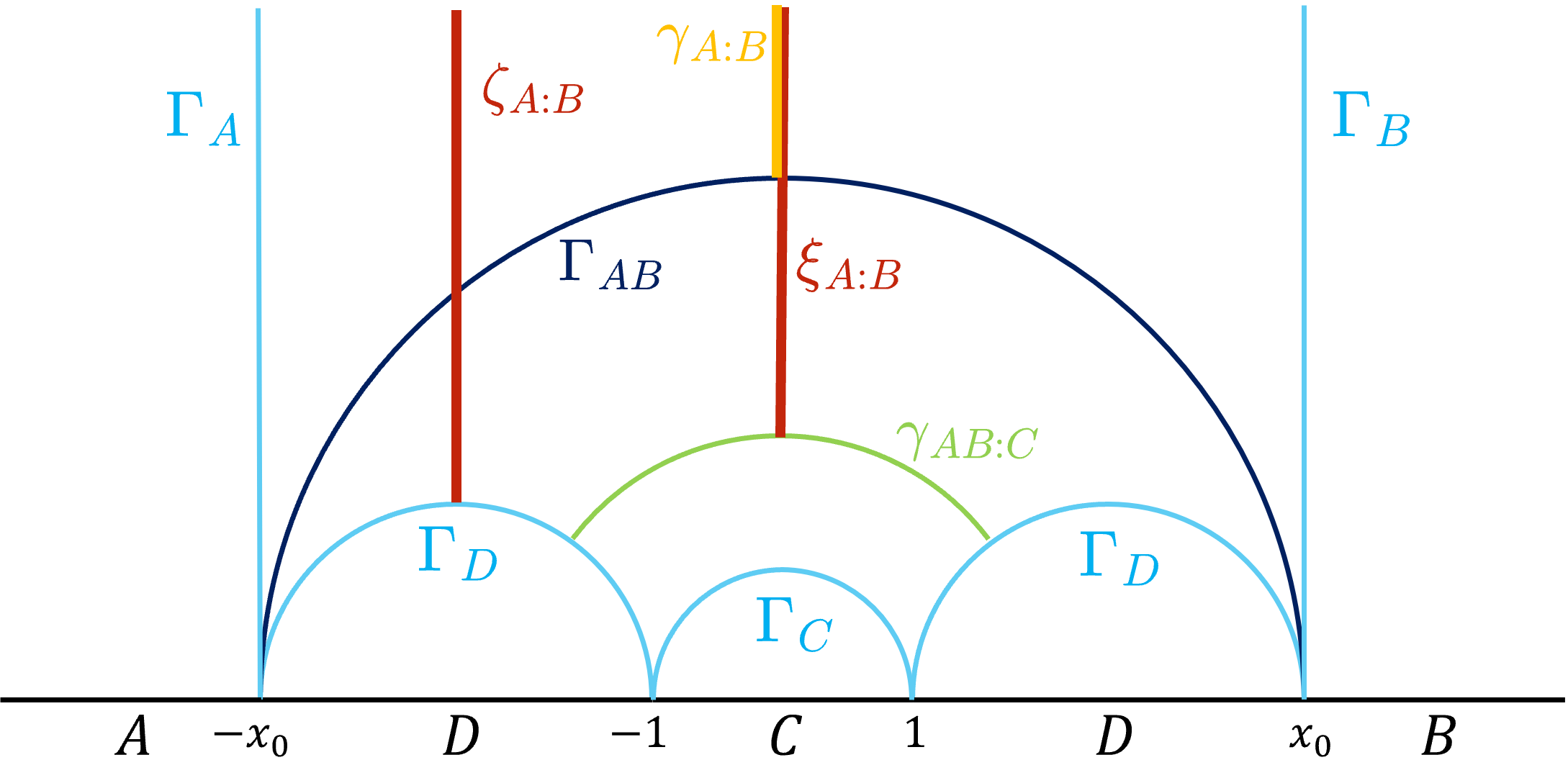}
    \caption{An upper half plane  example in which $Q_4$ and $\widetilde{Q}_4$ are not bounded in terms of $I_3$. The RT surfaces for each boundary subregion are shown in blue, in addition to $\gamma_{A:B}$ (green), $\xi_{A:B}$ (red), and $\zeta_{A:B}$ (yellow).}
    \label{fig:counter-example}
\end{figure}

Now, we show that for holographic states $\widetilde{Q}_4=Q_4=0$ in the interior of parameter regions within which $I_3=0$. 
To show this, consider the proof that $-I_3 \geq 0$ for any four regions $A$, $B$, $C$ and $D$ partitioning a state with a holographic dual \cite{Hayden:2011ag}. That proof involved taking the RT surfaces $\Gamma_{AB}$, $\Gamma_{AC}$, and $\Gamma_{BC}$ and recombining them into candidate surfaces for $\Gamma_A$, $\Gamma_B$, $\Gamma_C$, and $\Gamma_D$. 
In the interior of a parameter region in which  $I_3$ vanishes,  the candidate surfaces obtained from  $\Gamma_{AB}$, $\Gamma_{AC}$ and $\Gamma_{BC}$ are the same as the analogous RT surfaces for the single subregions. This implies
\begin{equation} \label{eqsurfaces}
    \Gamma_A \cup \Gamma_B \cup \Gamma_C \cup \Gamma_D = \Gamma_{AB} \cup \Gamma_{AC} \cup \Gamma_{BC}.  
\end{equation}
%\CM{I have changed the following to say endpoint/boundary because we don't really have endpoints in higher dimensional settings.}
The endpoints/boundaries of the cross-section $\gamma_{A:B}$ in the interior of the spacetime have to lie on $\Gamma_{AB}$. 
By entanglement wedge nesting in the canonical purification,  $\gamma_{A:B}$ lies between $EW(A)$ and $EW(B)$, so it cannot end on $\Gamma_A$ or $\Gamma_B$. 
Thus, $\gamma_{A:B}$ can only end on segments of $\Gamma_{AB}$ that are contained in $\Gamma_C \cup \Gamma_D$ on the left side of \eqref{eqsurfaces}. 
Thus all endpoints/boundaries of $\gamma_{A:B}$ in the interior of the spacetime lie on $\Gamma_C$ or $\Gamma_D$, so $\gamma_{A:B}$ is a valid candidate for $\zeta_{A:B}$. 
Since we already showed that $\widetilde Q_4 \geq 0$, this implies that $\gamma_{A:B} = \zeta_{A:B}$ and hence $\widetilde{Q}_4$ vanishes. 
From \eqref{eq:Q4-tilde-geq-Q4} we have $\widetilde{Q}_4 \geq Q_4 \geq 0$, so $I_3=0$ also implies $Q_4=0$. 

The argument above need not apply at configurations on the boundary of parameter space regions with $I_3=0$ where the RT surfaces, or equivalently, the entanglement wedges undergo a phase transition.
%\CM{I don't think Hayden et al say anything about this phase transition or when does $I_3$ vanish, so this citation feels a bit awkward.}
At these configurations, the RT surfaces used to compute $I_3$ and $Q_4$ change discretely. These jumps can cause $Q_4$ to change discontinuously, and hence be ill-defined, even though the areas of RT surfaces that compute $I_3$ change continuously. 
In particular, after passing to regions with non-zero $I_3$, $Q_4$ cannot be bounded in terms of $I_3$ as we shall show next.  

As an explicit demonstration, consider the constant time slice in vacuum AdS$_3$ -- this is a hyperbolic disc, which we map to the upper-half plane by a conformal transformation. 
The upper-half plane has metric $\mathrm{d}s^2 = \frac{1}{z^2}( \mathrm{d}z^2 + \mathrm{d}x^2)$, where we have set the AdS length scale, $\ell_{\text{AdS}} = 1$.  
We consider a case where the boundary subregions $A,B,C$ each have a single connected component, while $D$ has two components, as shown in Fig.~\ref{fig:counter-example}. 
We assume that the $EW(D)$, $EW(AC)$, and $EW(BC)$ are all in the disconnected phase, which requires $3 < x_0 < (3 + 2\sqrt{2})$.
In this case, we find
\begin{subequations}    
\begin{align}
    -I_3(A:B:C) 
    &= \frac{1}{2G} \ln \frac{4x_0}{(x_0-1)^2} , \\
    Q_{4}(A:B) 
    &= \frac{1}{4G} \ln \sqrt{x_0} , \\
    \widetilde{Q}_{4}(A:B) 
    &= \frac{1}{4G} \ln \frac{2 x_0}{x_0-1} . 
    \label{eq:Q4-tilde-example}
\end{align}
\end{subequations}
As we approach $x_0 = (3 + 2\sqrt{2})$, the triple information $-I_3$ smoothly goes to zero while $Q_{4}$ and $\widetilde{Q}_4$ remain finite. 
By choosing $x_0$ to be sufficiently close to its transition value, we can make $-I_3$ small enough to violate any inequality of the form $Q_4 \leq - c_{Q} I_3$ or $\widetilde{Q}_4 \leq - c_{\widetilde{Q}} I_3$.
Thus, no such inequalities can be formulated that are satisfied by all time-reflection-symmetric holographic states.

In this case, $\widetilde{Q}_4 - Q_4$ goes to zero as we approach $x_0 = (3+2 \sqrt{2})$, and one might wonder whether this difference could be bounded by $-I_3$. However, if we interchange the roles of  regions $C$ and $D$, the entanglement wedge cross-section $\gamma_{AB:C}$ coincides with $\Gamma_{AB}$. Then, $\xi_{A:B} = \gamma_{A:B}$, which implies $Q_4=0$, while $\widetilde{Q}_4$ is still given by \eqref{eq:Q4-tilde-example}.
Thus in this case, as we approach $x_0 = (3+2 \sqrt{2})$, $-I_3$ goes to zero but $\widetilde{Q}_4 - Q_4$ does not. So $\widetilde{Q}_4 - Q_4$ also cannot be bounded by in terms of $-I_3$. 

As mentioned earlier, the significance of the value $x_0 = (3 + 2\sqrt{2})$ is that there is a phase transition there where $EW(D)$ becomes connected and $I_3$ vanishes. Above the transition, $EW(ABC) = EW(AB) \cup EW(C)$, which implies that both $I_3$ and $Q_4$ vanish by the result of \cite{Balasubramanian:2024ysu}. 
Since $I_3$ is built from entanglement entropies, it transitions to zero continuously. 
However, $Q_4$ and $\widetilde{Q}_4$ involve entanglement wedge cross-sections, and they change discontinuously. 

\section{Discussion}
\label{sec:discussion}

In this work, we have showed that, for pure time-reflection-symmetric holographic states, all currently known signals of four-party entanglement vanish if the triple information, $I_3$, vanishes.  These signals include all quantities built from the multi-entropy \cite{Gadde:2022cqi, Gadde:2026msg} and the residual information $Q_4$ (except on a measure zero set).  In this sense, $I_3$ is the strongest known signal of the presence of four-party entanglement in holographic states.

We also proved that $0 \leq \cM_{AB|CD} \leq -I_3$ and, that the genuine multi-entropy therefore obeys $0 \leq \GMo \leq -I_3$.
Evidently, in holographic systems, all four-party multi-entropy signals are bounded in terms of the triple information. 
After this paper appeared, these bounds have been further developed in terms of a holographic multi-entropy cone in \cite{Ju:2026zbu}, demonstrating that our bounds are tight.

In previous work \cite{Balasubramanian:2025hxg}, we similarly bounded the three-party genuine multi-entropy in terms of the residual information, $0 \leq GM^{(3)} \leq \frac{1}{2} R^{(3)}$.  
Thus, if one aims to understand the structure of four-party entanglement in holographic systems, most of the information is captured by $R^{(3)}$ and $I_3$ -- the genuine multi-entropies $GM^{(3)}$ and $GM^{(4)}$ are weaker signals. It would be interesting to explore whether this pattern continues for more  parties. 

Additionally, we also provide a counterexample to demonstrate that the residual entropy $Q_4$ cannot generally be bounded by the triple information.
Our counterexample suggests that $Q_4$ carries information that is not contained in $I_3$. 
Although $I_3$ goes to zero smoothly at the entanglement wedge phase transition, the behavior of $Q_4$ shows that a finite amount of four-party entanglement can persist up to this point. 
This indicates that the entanglement structure of the state undergoes a discontinuous change at such transitions. 
Similar discontinuities were  discussed in a time-dependent context in \cite{Balasubramanian:2025jhq,Balasubramanian:2011at}, and understanding  the underlying mechanism, for example by building simple models where similar features appear, is an important direction for future work. 

The bounds found here and in \cite{Balasubramanian:2025hxg} are valid only in holography.  For example, they are explicitly violated by pure states that are  GHZ-like. 
Overall, our results reveal new  constraints on quantum states with geometric dual descriptions, the complete characterization of which  remains an important open question.
%of the entanglement structure of holographic states remains an important open question. Our results give new constraints; understanding how constraining they are is a key goal for the future.

\section{Acknowledgements}
\noindent MJK  thanks the organizers and participants of the conference \emph{Southwest Strings Meeting 2026} held in Arizona State University, Tempe AZ, for the stimulating environment and discussions. 
VB, WKLM, and CM were supported in part by the DOE through DE-SC0013528 and the QuantISED grant DE-SC0020360.   MJK is supported by the DOE through DE-SC0010813 and the Start-up Research Grant for new faculty provided by Texas A\&M University.  SFR is supported in part by STFC through grant number ST/X000591/1. 
%VB thanks the International Center for Theoretical Sciences (ICTS), Bengaluru for support through the Asha Gupta Chair Professorship while this paper was being completed.

\bibliography{main}

\end{document}